\documentclass[pdftex,twocolumn,epjc3]{svjour3}          

\RequirePackage[T1]{fontenc}

\smartqed  

\RequirePackage{graphicx}
\RequirePackage{mathptmx}      
\RequirePackage{flushend}
\RequirePackage[numbers,sort&compress]{natbib}
\RequirePackage[colorlinks,citecolor=blue,urlcolor=blue,linkcolor=blue]{hyperref}

\journalname{Eur. Phys. J. C}

\def\beq{\begin{equation}}
\def\eeq{\end{equation}}
\def\bea{\begin{eqnarray}}
\def\eea{\end{eqnarray}}
\def\beal{\begin{align}}
\def\eeal{\end{align}}
\def\sbeq{\begin{subequations}}
\def\seeq{\end{subequations}}

\def\na{{\bf \nabla}}
\def\pa{\partial}

\def\nn{\nonumber}

\begin{document}

\title{ The Influence of the Shear on the Gravitational Waves in the Early Anisotropic Universe}


\author{Yoo Geun Song\thanksref{e1,addr1, addr2}
        \and
        Seok Jae Park\thanksref{e2,addr1,addr2} 
}

\thankstext[$\star$]{t1}{Thanks to the title}
\thankstext{e1}{e-mail: ygsong@kasi.re.kr}
\thankstext{e2}{e-mail: sjpark@kasi.re.kr}

\institute{Korea Astronomy and Space Science Institute, Daejeon 305-348, Korea\label{addr1}
          \and
          University of Science and Technology,  Daejeon 305-350, Korea\label{addr2}
}

\date{Received: date / Accepted: date}

\maketitle

\begin{abstract}

We study the singularity of the congruences for both timelike and null geodesic curves using the expansion of the early anisotropic Bianchi type I Universe.  In this paper, we concentrate on the influence of the shear of the timelike and null geodesic congruences in the early Universe. Under some natural conditions, we derive the Raychaudhuri type equation for the expansion and the shear-related equations.
Recently, scientists working on the LIGO (Laser Interferometer Gravitational-Wave Observatory) have shown many possibilities to observing the anisotropy of the ``primordial'' gravitational wave background radiation. We deduce the evolution equation for the shear that may be responsible for those observational results.

\end{abstract}

\section{Introduction}
\setcounter{equation}{0}
\renewcommand{\theequation}{{3}.\arabic{equation}}
The existence of Hawking-Penrose (HP) singularity~\\
\cite{HP70} can be assumed at the beginning of the Universe that was born with the Big Bang as an unimaginary hot, dense point.
If we presume that the early Universe was filled with a perfect fluid containing massive particles and/or massless particles,
then we could find equations of state for each particle, by using the strong energy condition that was used to show the HP singularity theorem.

The motion types of timelike and null geodesic congruences in the Universe can be described in terms of the expansion, the shear, and the rotation.
In this article, we will describe the expansion rates of
timelike and null geodesic congruences and the shearing motions in the early anisotropic Universe. The aspects of the rotational motions are negligible to produce the Bianchi type I Universe.
Additionally, we derive the Raychaudhuri type equation that is an evolution equation for the expansion.

There are many results regarding the dynamics of geodesic surface congruence in the early Universe ~\\\cite{CH2007,CH2008,CH2011}, \\
\cite{CH2011Book}.
In this article, we will going to discuss the homogeneous and anisotropic Universe. In particular, we consider the well-known models of the homogeneous and anisotropic Universe such as Bianchi type I and Kasner Universe models ~\cite{Kasner21}.
The Kasner Universe is the ``vacuum'' Bianchi type I Universe.
We also suggest a straightforward approach to investigate the Universe models.
Many scientists have studied the Bianchi type I Universe (e.g., \cite{BS98,Caceres2010, Chiba97, CS95,Jacobs69,Tsagas2008, PM2015,SB2015, ABDR2011, AGKB2010, Tiwari2008, Saha2007, PM2015, SZ2010, SS2015, SK2011, SB2015, Shamir2015, JAMM2012}).

This article is organized as follows.
In Section 2, we discuss the kinematical quantities in the Bianchi type I Universe and introduce the formalism that describes the geodesic congruence in the early Universe.
In Section 3, we describe the timelike geodesic curves in the Bianchi type I Universe.
We
investigate the expansion rate of timelike geodesic congruence and the aspects of shearing motions in the Bianchi type I Universe model.
In Section 4, we describe the null geodesic curves in the Bianchi type I Universe.
We investigate the expansion rate of null geodesic congruence and the aspects of the shearing motions, in our Universe model, for null case.
Finally, we conclude in Section 5.

\section{The Kinematical Quantities in the Bianchi Type I Universe}
\setcounter{equation}{0}
\renewcommand{\theequation}{\arabic{section}.\arabic{equation}}


In this section, we review the research works by Cho and Hong (e.g., \cite{CH2007,CH2008,CH2011,CH2011Book}) on the expansion, the shear, and the rotation which will be described later.
 We also discuss the well-known models of homogeneous and anisotropic Universe such as Bianchi type I and Kasner Universe models 
  (e.g., \cite{Ellis2009, WE97, HE73,EV99}).

In order to define the action on the curved space-time manifold, we let $(M,
g_{ab})$ be a $4$-dimensional manifold associated with the metric
$g_{ab}$.  Given $g_{ab}$, we can have a unique covariant
derivative $\na_{a}$ satisfying~\cite{Wald84} \bea
\na_{a}g_{bc}&=&0,\nn\\
\na_{a}\omega^{b}&=&\pa_{a}\omega^{b}+\Gamma^{b}_{~ac}~\omega^{c},\nn\\
(\na_{a}\na_{b}-\na_{b}\na_{a})\omega_{c}&=&R_{abc}^{~~~d}~\omega_{d}.\label{rtensor}
\eea

We parametrize the surface generated by the world sheet coordinate $t$, and then we
have the corresponding vector field $\xi^{a}=(\pa/\pa t)^{a}$. Then $\xi^a$ satisfies the timelike condition 
$\xi\cdot\xi=-1$. We introduce the tensor field $B_{ab}$ defined by
 \beq
  B_{ab}=\na_{b}\xi_{a}, \label{babbbabt}
  \eeq
 which satisfies the following identity
 \beq
 B_{ab}\xi^{a}=0.
 \eeq 
Next, we introduce the metric $h_{ab}$, \beq
h_{ab}=g_{ab}+\xi_{a}\xi_{b},
\label{projections}\eeq
which satisfies
\bea
h_{ab}\xi^{a}=0,~~~~~ h_{ab}\xi^{b}=0,~~~~~
h_{ab}g^{bc}h_{cd}=h_{ad},~~~~~ h_{ab}h^{ab} = 3.
\eea
Here note that $h_{ab}$ is the metric on the
hypersurfaces orthogonal to $\xi^{a}$. Moreover, we can define projection operator
$h^{a}_{~b}$ as \beq
h^{a}_{~b}=g^{ac}h_{cb}.\eeq
This operator fulfills \beq \begin{array}{cc}
h^{a}_{~b}h^{b}_{~c}=h^{ab}h_{bc}=h^{a}_{~c} ~,~ ~h_{ab}h^{bc}h_{cd}=h_{ad}.
\end{array}\eeq

Now, we decompose $B_{ab}$ into three pieces \beq
B_{ab}=\frac{1}{3}\theta h_{ab}+\sigma_{ab}+\omega_{ab}.\label{bab}\eeq
In the above equation, we defined kinematical quantities of the
timelike geodesic congruence in the Bianchi type I Universe, such that, the expansion $\theta$, the shear $\sigma_{ab}$, and the rotation $\omega_{ab}$, that are
given by
\beq\theta=B^{ab}h_{ab},~~~
\sigma_{ab}=B_{(ab)}-\frac{1}{3}\theta h_{ab},~~~
\omega_{ab}=B_{[ab]}.\label{thetasigma}\eeq
We then find
\bea
\sigma_{ab}h^{ab}=0, ~~~~~\omega_{ab}h^{ab}=0,
\eea and
\beq \xi^{c}\na_{c}B_{ab}
=-B^{c}_{~b}B_{ac}+R_{cbad} \xi^{c}\xi^{d}.
\label{bbr}\eeq

Exploiting Equation (\ref{bbr}), one arrives at \bea
\xi^{a}\na_{a}\theta
&=&-\frac{1}{3}\theta^{2}-\sigma_{ab}\sigma^{ab}
+\omega_{ab}\omega^{ab}-R_{ab}\xi^{a}\xi^{b},\label{eq1}\\
\xi^{c}\na_{c}\omega_{ab}&=&-\frac{2}{3}\theta(\omega_{ab}-\xi^{c}\xi_{[a}\omega_{b]c})
-2\sigma^{c}_{~[b}\omega_{a]c},\label{eq2}\\
\xi^{c}\na_{c}\sigma_{ab}&=&-\frac{1}{9}\theta^{2}\xi_{a}\xi_{b}
-\frac{2}{3}\theta h^{c}_{~(a}\sigma_{b)c}-\sigma_{ac}\sigma^{c}_{~b}
-\omega_{ac}\omega^{c}_{~b} \nn\\
&&+\left(R_{c(ab)d}+\frac{1}{3}g_{ab}R_{cd}\right)\xi^{a}\xi^{b} +\frac{1}{3}g_{ab}(\sigma_{cd}\sigma^{cd}-\omega_{cd}\omega^{cd}) \nn\\
&&-\frac{1}{3}\theta\xi^{c}\xi_{(a}\na_{|c|}\xi_{b)}-\frac{1}{3}\xi_{a}\xi_{b}\xi^{c}\na_{c}\theta
.
\label{eq3}
\eea

We consider the comoving coordinates $(t, x, y, z)$ such that the metric takes the form
\beq
ds^2 = - dt^2 + X^2 (t) dx^2 + Y^2 (t) dy^2 + Z^2 (t) dz^2 ~.
\eeq
We redefine the ``scale factor'' $l(t)$ by $l^3 = XYZ$.
Here, $X(t)$, $Y(t)$, and $Z(t)$ are the scale factors of $x-$, $y-$, and $z-$ directions, respectively.
If $X=Y=Z$, we get the usual Friedmann-Robertson-Walker (FRW) space-time.

Now, we consider the Kasner vacuum Universe -- the ``vacuum'' case of the Bianchi type I Universe -- which satisfies $\rho = 0$ and $P = 0$, where $\rho$ and
$P$ are the mass-energy density and pressure of the fluid as
measured in its rest frame, respectively~\cite{Wald84,MTW}. Then, $X$, $Y$, and $Z$ become distinct functions of $t$. 
The metric and the initial conditions of the Kasner vacuum Universe, except $\rho$ and $P$, are exactly the same as those of the Bianchi type I Universe.
Now, we set
\beq
\xi^a \na_a \xi^b = 0 ~,~~ \omega_{ab} = 0 ~,~~ B_{(ab)}=\theta_{ab} \ne 0~, ~~\sigma_{ab} \ne 0 ~.
\eeq
The meaningful results of the field equations are  \cite{EV99}
\bea
\frac{3 \ddot{l}}{l} + 2 \sigma ^2 = 0 ~, ~~~
\left(l^6 \sigma^2 \right)^{\cdot} = 0 ~,\label{fieldeq1}
\eea
where
\beq
\dot{l} = \xi^a\na_a l ~~,~~ \sigma^2 \equiv \frac{1}{2} \sigma_{ab}\sigma^{ab} = \frac{\Sigma^2}{l^6} ~. 
\eeq
From Equations (\ref{fieldeq1}), we have
\beq
\sigma_{ab} = \frac{\Sigma_{ab}}{l^3} ~,~ ~2~ \Sigma^2 = (\Sigma_{11})^2 + (\Sigma_{22})^2 + (\Sigma_{33})^2 ~.
\eeq
In this case, from Equations (\ref{thetasigma}), in orthonormal bases, the symmetric part $B_{(ab)}$ of $B_{ab}$ is given by
\beq
B_{(ab)} = \sigma_{ab} + \frac{1}{3} \theta h_{ab} = \frac{\Sigma_{ab}}{l^3} + \frac{\dot{l}}{l} \delta_{ab} ~, \eeq
where \beq\theta = \frac{3\dot{l}}{l}~.\eeq
Then, we can obtain
\bea
\frac{\dot{X}}{X} = \frac{\Sigma_{11}}{l^3} + \frac{\dot{l}}{l} ~~,~~\frac{\dot{Y}}{Y} = \frac{\Sigma_{22}}{l^3} + \frac{\dot{l}}{l} ~~,~~ \frac{\dot{Z}}{Z} = \frac{\Sigma_{33}}{l^3} + \frac{\dot{l}}{l} ~.
\label{xxyyzz}
\eea
We can also obtain
\bea
\Sigma_{11} &=& \frac{2}{\sqrt{3}} ~\Sigma~ \sin \alpha ~, \nn\\ 
\Sigma_{22} &=& \frac{2}{\sqrt{3}} ~\Sigma~ \sin \left(\alpha  + \frac{2}{3} \pi \right) ~,\nn\\
\Sigma_{33} &=& \frac{2}{\sqrt{3}} ~\Sigma~ \sin \left(\alpha  + \frac{4}{3} \pi \right)  ~,\label{sigmas}
\eea
where $\alpha ~\left(-{\pi}/{6} < \alpha \le {\pi}/{2}\right)$ is a constant determining the direction in which the most rapid expansion takes place.

%
%
%

Let us consider the time-reverse of the model.
For general values of $\alpha$, i.e., $-{\pi}/{6} < \alpha < {\pi}/{2}$, the term $1 + 2 \sin(\alpha + ({4 \pi}/{3}))$ will be negative.
Thus, if we consider the forward direction of time, we have a ``cigar'' singularity: matter collapses in along the $z-$axis from infinity, halts, and then starts re-expanding, while in the $x-$ and $y-$directions the matter expands at all times.
In the exceptional case (i.e., $\alpha = {\pi}/{2}$), the terms $1 + 2 \sin(\alpha + ({2 \pi}/{3}))$ and $1 + 2 \sin(\alpha + ({4 \pi}/{3}))$ both vanish.
Then we have a ``pancake'' singularity: matter expands in all directions, starting from an indefinitely high expansion rate in the $x-$direction but from zero expansion rates in the $y-$ and $z-$directions ~ 
(e.g., \cite{Ellis2009, WE97, HE73}).

\medskip

In the ``non-vacuum'' (i.e., $\rho \ne 0$ and $P \ne 0$) anisotropic Bianchi type I Universe, we can assume the perfect fluid.
For a perfect fluid, the energy-momentum tensor is given by
\beq
T_{ab}=\rho~\xi_{a}\xi_{b}+P~(g_{ab}+\xi_{a}\xi_{b}) ~.
\eeq

\bigskip

\section{Timelike Geodesic Curves in the Bianchi Type I Universe}
\setcounter{equation}{0}
\renewcommand{\theequation}{\arabic{section}.\arabic{equation}}

The motion types of timelike geodesic congruence in the Universe can be described in terms of the expansion, the shear, and the rotation.
In this section, we describe the expansion rate of timelike geodesic congruence and the shearing motions in the early anisotropic Universe.

In the Bianchi type I space-time, Equations (\ref{eq1})-(\ref{eq3}) become
\bea
\xi^{a}\na_{a}\theta
&=&-\frac{1}{3}\theta^{2}-\sigma_{ab}\sigma^{ab}
-R_{ab}\xi^{a}\xi^{b},\label{neweq1}\\
\xi^{c}\na_{c}\omega_{ab}&=&0~,\label{neweq2}\\
\xi^{c}\na_{c}\sigma_{ab}&=&-\frac{1}{9}\theta^{2}\xi_{a}\xi_{b}
-\frac{2}{3}\theta h^{c}_{~(a}\sigma_{b)c}-\sigma_{ac}\sigma^{c}_{~b} \nn\\
&& +\left(R_{c(ab)d} +\frac{1}{3}g_{ab}R_{cd}\right)\xi^{a}\xi^{b}+\frac{1}{3}g_{ab}\sigma_{cd}\sigma^{cd} \nn\\
&&-\frac{1}{3}\theta\xi^{c}\xi_{(a}\na_{|c|}\xi_{b)}-\frac{1}{3}\xi_{a}\xi_{b}\xi^{c}\na_{c}\theta
.
\label{neweq3}
\eea
Since,
\beq
\xi^{a}\na_{a}\theta = \dot{\theta} ~,
\eeq
and from Equation (\ref{neweq1}), we can obtain a Raychaudhuri type equation
\bea
\dot{\theta}&=&-\frac{1}{3}\theta^{2}
-\sigma_{ab}\sigma^{ab}
-R_{ab}\xi^{a}\xi^{b}\nn\\
&=&-\frac{3 \dot{l}^2}{l^2}
-2\frac{\Sigma^2}{l^6}
-R_{ab}\xi^{a}\xi^{b} ~, ~\label{rteq}
\eea
by using the fact that $\theta = {3 \dot{l}}/{l}$ and $\sigma_{ab} = {\Sigma_{ab}}/{l^3}$.
We now assume that the strong energy condition
\beq
R_{ab}\xi^{a}\xi^{b}=8\pi\left(T_{ab}
-\frac{1}{2}Tg_{ab}\right)\xi^{a}\xi^{b} \ge 0 ~.\label{strong}\eeq
where $T_{ab}$ and $T$ are the energy-momentum
tensor and its trace, respectively.
If $R_{ab}\xi^{a}\xi^{b}= 0$, then this condition corresponds to the strong energy condition for the Kasner ``vacuum'' (i.e., $T_{ab} = 0$ and $T = 0$) Universe.
The Raychaudhuri type equation (\ref{rteq}) then has a solution of the form
\beq
\frac{1}{\theta(\tau)}~\ge~
\frac{1}{\theta(0)}+\frac{1}{3} \left[\tau + \frac{2}{3} \Sigma^2 \int_{0}^{\tau} \frac{1}{l^4 ~ \dot{l}^2} ~ d t\right]~,
\label{solnofrteq}
\eeq
where $\tau$ is the proper time.
We assume that $\theta(0)$ is negative so that the
congruence is initially converging as in the point particle case.
The inequality (\ref{solnofrteq}) implies that $\theta(\tau)$ must pass
through the singularity within a proper time (see Figure 2),
\beq
\tau \le \frac{3}{|\theta(0)|}- \frac{2}{3} \Sigma^2 \int_{0}^{\tau} \frac{1}{l^4 ~ \dot{l}^2} ~ {d} t~,\label{propertime}
\eeq
since,
\beq
 \frac{3}{|\theta(0)|} > 0~.
 \eeq
If we choose $\tau > 0$ such that,
\beq \frac{3}{|\theta(0)|} \ge \frac{2}{3} \Sigma^2 \int_{0}^{\tau} \frac{1}{l^4 ~ \dot{l}^2} ~ {d} t~ , \eeq
then the right-hand side of Equation (\ref{propertime}) is greater than or equal to zero.

For a perfect fluid, the strong energy condition (\ref{strong}) yields one inequality equation \beq 4\pi (\rho + 3 P) \ge 0.
\label{sectimelike} \eeq
Then we have the following two inequalities
\beq
\rho+3P\ge0,~~~\rho+P\ge0.\eeq
If we neglect the shearing motions (i.e., $\sigma_{ab} = 0$), then we have the
differential inequality equation
\beq
\frac{d\theta}{d\tau}+\frac{1}{3}\theta^{2}\le 0 ~,
\eeq which has a
solution in the following form
\beq\frac{1}{\theta(\tau)}\ge
\frac{1}{\theta(0)}+\frac{1}{3}\tau.\eeq
If we assume that $\theta(0)$ is negative, then the expansion $\theta(\tau)$ must go to the
negative infinity along that geodesic within a proper time
\beq\tau\le \frac{3}{|\theta(0)|},\eeq  whose consequence coincides with
that of Hawking and Penrose \cite{HP70}.


\bigskip

From now on, we will consider the shear of timelike geodesic
congruence in the early anisotropic Universe.
Using Equation (\ref{neweq3}), we obtain an evolution equation for the shear,
\bea
\frac{d\sigma_{ab}}{d t}
&=&-\frac{1}{9}\theta^{2}\xi_{a}\xi_{b}-\frac{2}{3}\left[\theta \left(\frac{1}{l^3} h^{c}_{~(a}\Sigma_{b)c}\right)\right]\nn\\
&&- \frac{1}{l^6} \Sigma_{ac}\Sigma^{c}_{~b}
+\left(R_{c(ab)d}+\frac{1}{3}g_{ab}R_{cd}\right)\xi^{a}\xi^{b}\nn\\
&&+\frac{2}{3 l^6}g_{ab}\Sigma^2 -\frac{1}{3}\theta\xi^{c}\xi_{(a}\na_{|c|}\xi_{b)} \nn\\
&&-\frac{1}{3}\xi_{a}\xi_{b}\xi^{c}\na_{c}\theta ~. \label{sigma1}
\eea

Substituting Equation (\ref{rteq}) into (\ref{sigma1}), we obtain
\bea
\frac{d\sigma_{ab}}{d t}
&=&-\frac{2 \dot{l}}{l^4} h^{c}_{~(a}\Sigma_{b)c}\nn\\
&&- \frac{1}{l^6} \Sigma_{ac}\Sigma^{c}_{~b}
+R_{c(ab)d} \xi^{a}\xi^{b}\nn\\
&&+\frac{2 ~ \Sigma^2}{3 ~ l^6}~ h_{ab} -\frac{\dot{l}}{l}\xi^{c}\xi_{(a}\na_{|c|}\xi_{b)} ~.
\eea

In the standard point-particle inflationary cosmology,
the influence of the shear on the ensuing Universe
evolution are negligible to produce the homogeneous and isotropic Universe features.
It is worthy to note that in the homogeneous and ``anisotropic'' Universe, one can have the condition
$\sigma_{ab} \neq 0$. Here the non-vanishing
$\sigma_{ab}$ evolves and dominates in the early anisotropic Universe.

We assume the following condition to investigate the evolution of shear in the ``extremely early'' Universe (e.g., in the inflationary epoch),
\beq
\frac{d\sigma_{ab}}{d t}
\approx - \frac{1}{l^6} \Sigma_{ac}\Sigma^{c}_{~b}
+\frac{2 ~ \Sigma^2}{3 ~ l^6}~h_{ab} ~,\label{sigmaab}
\eeq
where we left only two $l^{-6}$ terms.
If we {sub}stitute Equations (\ref{sigmas}) into (\ref{sigmaab}), then we finally get
\bea
\frac{d\sigma_{ab}}{d t}
 &\approx& - \frac{2 ~ \Sigma^2}{l^6}~\left(- 1 + \frac{1}{X^2} + \frac{1}{Y^2} + \frac{1}{Z^2} \right) ~.\label{finaleq}
\eea
By integrating Equation (\ref{finaleq}), we have the following approximation on the shear,
\bea
\sigma_{ab} (\tau) 
 &\approx& \sigma_{ab} (0)  - ~2 ~ \Sigma^2~
 \int_{0}^{\tau} \frac{1}{l^6} \left( {\rm tr}  (g^{ab} ) \right) ~{d} t ~.
\eea
Suppose one considers the time-reverse of the model, then the evolution equation for the shear depends on the scale factor $l(t)$ in the ``extremely early'' anisotropic Universe.

\bigskip
\bigskip

\section{Null Geodesic Curves in the Bianchi Type I Universe}
\setcounter{equation}{0}
\renewcommand{\theequation}{\arabic{section}.\arabic{equation}}


In this section, we
consider the evolution of the vectors in a \\ $2-$dimensional
subspace of spatial vectors normal to the null tangent vector
field $k^{a}=(\pa/\pa\lambda)^{a}$, where $\lambda$ is the affine
parameter, and to an auxiliary null vector $l^{a}$ which points in
the opposite spatial direction to $k^{a}$, normalized by~\cite{Carroll2004}
\beq
l^{a}k_{a}=-1~,
\eeq and is parallel transported,
namely,
\beq
k^{a}\na_{a}l^{b}=0.
\eeq
The spatial vectors in the
$2-$dimensional subspace are then orthogonal to both $k^{a}$
and $l^{a}$.
We now introduce the metric $n_{ab}$, \beq
n_{ab}=g_{ab}+k_{a}l_{b}+l_{a}k_{b}.
\label{projection2}\eeq
Similar to the timelike case, we introduce tensor fields
\beq
B_{ab}=\na_{b}k_{a},
\eeq
satisfying the identity
\beq
B_{ab}k^{a}=0.
\eeq

We decompose $B_{ab}$ into three pieces \beq
B_{ab}=\frac{1}{2}\theta
n_{ab}+\sigma_{ab}+\omega_{ab},\label{babnull2}\eeq where the
expansion, shear, and rotation of the null geodesic congruence along the
affine direction are defined as
\beq
\theta=B^{ab}n_{ab},~~~
\sigma_{ab}=B_{(ab)}-\frac{1}{2}\theta n_{ab},~~~
\omega_{ab}=B_{[ab]}.
\eeq
The metric $n_{ab}$ also satisfies the identities
\beq
\sigma_{ab}n^{ab}=\omega_{ab}n^{ab}=0,
\eeq
and
\bea
&&n_{ab}k^{a}=n_{ab}k^{b}=n_{ab}l^{a}=n_{ab}l^{b}=0,~~~\nn\\
&&n_{ab}g^{bc}n_{cd}=n_{ad},~~~
n_{ab}n^{ab}=2.
\eea
We define $n^{a}_{~b}$ as
\beq
n^{a}_{~b}=g^{ac}n_{cb}=\delta^{a}_{~b}+k^{a}l_{b}+l^{a}k_{b},
\eeq
which fulfills the following identities
\beq
k^{c}\na_{c}n^{a}_{~b}=0,
\eeq
and
\bea
&&n^{a}_{~b}k^{b}=n^{a}_{~b}k_{a}=n^{a}_{~b}l^{b}=n^{a}_{~b}l_{a}=0,~~~\nn\\
&&n^{a}_{~b}n^{b}_{~c}=n^{a}_{~c},~~~
n_{ab}n^{ac}=n_{b}^{~c}, \nn\\
&& n_{a}^{~b}n_{bc}=n_{ac}.
\eea
Then, we have the identities
\beq
B_{ab}k^{a}=0,~~~\sigma_{ab}k^{b}=0,~~~
\omega_{ab}k^{b}=0,
\eeq
and
\beq
k^{c}\na_{c}B_{ab}=-B^{c}_{~b}B_{ac}+R_{cbad}k^{c}k^{d}.
\label{bbrnull}\eeq

Using Equation (\ref{bbrnull}), we get
\bea
k^{a}\na_{a}\theta
&=&-\frac{1}{2}\theta^{2}-\sigma_{ab}\sigma^{ab}+\omega_{ab}\omega^{ab}-R_{ab}k^{a}k^{b},\label{eq1null}\\
k^{c}\na_{c}\omega_{ab}
&=&-\theta(\omega_{ab}-k^{c}k_{[a}\omega_{b]c})
-2\sigma^{c}_{~[b}\omega_{a]c},\label{eq2null}\\
k^{c}\na_{c}\sigma_{ab}
&=&-\frac{1}{4}\theta^{2}k_{a}k_{b}
-\theta h^{c}_{~(a}\sigma_{b)c}
-\sigma_{ac}\sigma^{c}_{~b}-\omega_{ac}\omega^{c}_{~b}
\nn\\
&&+\left(R_{c(ab)d}+\frac{1}{2}g_{ab}R_{cd}\right)k^{c}k^{d}\nn\\
&&+\frac{1}{2}g_{ab}(\sigma_{cd}\sigma^{cd}-\omega_{cd}\omega^{cd})-\frac{1}{2}\theta k^{c}k_{(a}\na_{|c|}k_{b)}\nn\\
&&-\frac{1}{2}k_{a}k_{b}k^{c}\na_{c}\theta.\label{eq3null}
\eea
In the Bianchi type I space-time, Equations (\ref{eq1null})-(\ref{eq3null}) become
\bea
k^{a}\na_{a}\theta
&=&-\frac{1}{2}\theta^{2}-\sigma_{ab}\sigma^{ab}-R_{ab}k^{a}k^{b},\label{neq1null}\\
k^{c}\na_{c}\omega_{ab}
&=&0,\label{neq2null}\\
k^{c}\na_{c}\sigma_{ab}
&=&-\frac{1}{4}\theta^{2}k_{a}k_{b}
-\theta h^{c}_{~(a}\sigma_{b)c}
-\sigma_{ac}\sigma^{c}_{~b}
\nn\\
&&+\left(R_{c(ab)d}+\frac{1}{2}g_{ab}R_{cd}\right)k^{c}k^{d}\nn\\
&&+\frac{1}{2}g_{ab}\sigma_{cd}\sigma^{cd}-\frac{1}{2}\theta k^{c}k_{(a}\na_{|c|}k_{b)}\nn\\
&&-\frac{1}{2}k_{a}k_{b}k^{c}\na_{c}\theta ~,\label{neq3null}
\eea
since,
 \beq
  k^{a}\na_{a}\theta = \frac{d \theta}{d \lambda} ~.\eeq
From Equation (\ref{neq1null}), we can obtain the  Raychaudhuri type equation for null case,
\bea
\frac{d\theta}{d \lambda}&=&-\frac{1}{2}\theta^{2}-\sigma_{ab}\sigma^{ab}-R_{ab}k^{a}k^{b}\nn\\
&=&-\frac{9}{2} \left(\frac{\dot{l}}{l}\right)^2
-2\frac{\Sigma^2}{l^6}
-R_{ab}k^{a}k^{b}~, ~\label{rteqnull}
\eea
which resembles Equation (\ref{rteq}).

By Einstein's equation, we obtain
\beq
R_{ab}k^{a}k^{b}=8\pi ~ T_{ab} ~ k^{a}k^{b}  ~,\label{strongnull}\eeq
where $T_{ab}$ is the energy-momentum tensor.
Using the laws of physics, we may assume that the energy density $T_{ab} \xi^a \xi^b \ge 0$ for timelike case.
By continuity, we may also assume $T_{ab} k^a k^b \ge 0$ for null case.

The Raychaudhuri type equation (\ref{rteqnull}) for null case then has a solution of the form
\beq
\frac{1}{\theta(\lambda)}~\ge~
\frac{1}{\theta(0)}+\frac{1}{2} \left[\lambda + \frac{4}{9} \Sigma^2 \int_{0}^{\lambda} \frac{1}{l^4 ~ \dot{l}^2} ~ d \lambda \right] ~,
\label{solnnull}
\eeq
where $\theta(0)$ is the initial value of $\theta$ at
$\lambda=0$.
We assume again that $\theta(0)$ is negative. The
inequality (\ref{solnnull}) then implies that $\theta$ must pass
through the singularity within an affine length~
\beq \lambda\le
\frac{2}{|\theta(0)|}-\frac{4}{9} \Sigma^2 \int_{0}^{\lambda} \frac{1}{l^4 ~ \dot{l}^2} ~ d \lambda \le
\frac{2}{|\theta(0)|} ~.
\label{propertimenull} \eeq

Using Equation  (\ref{neq3null}), we obtain an evolution equation for the shear of null geodesic congruence,
\bea
\frac{d\sigma_{ab}}{d \lambda}
&=&-\frac{9}{4} \left(\frac{\dot{l}}{l}\right)^2 k_{a}k_{b} - 3 \frac{\dot{l}}{l^4} h^c_{(a} \Sigma_{b)c}
\nn\\
&&+\left(R_{c(ab)d}+\frac{1}{2}g_{ab}R_{cd}\right)k^{c}k^{d}\nn\\
&&-\frac{1}{l^6} \Sigma_{ac} \Sigma^{c}_{~b}+\frac{1}{l^6} g_{ab}\Sigma^2\nn\\
&&-\frac{3 \dot{l}}{2 l} k^{c}k_{(a}\na_{|c|}k_{b)}
-\frac{1}{2}k_{a}k_{b}k^{c}\na_{c}\theta ~,\label{sheareqnull}
\eea
where $k^{c}\na_{c}\theta = {d \theta}/{d \lambda}$.
Substituting Equation (\ref{rteqnull}) into (\ref{sheareqnull}), we obtain
\bea
\frac{d\sigma_{ab}}{d \lambda}
&=& - \frac{3 \dot{l}}{l^4} h^c_{(a} \Sigma_{b)c}
+\left(R_{c(ab)d}+\frac{1}{2}g_{ab}R_{cd}\right)k^{c}k^{d}\nn\\
&&-\frac{1}{l^6} \Sigma_{ac} \Sigma^{c}_{~b}+\frac{1}{l^6} \Sigma^2 \left(g_{ab} + k_a k_b\right)\nn\\
&&-\frac{3 \dot{l}}{2 l} k^{c}k_{(a}\na_{|c|}k_{b)}
~.
\eea

Since the expansion and the curvature are reciprocal, we may assume the following condition to investigate the evolution of shear in the ``extremely early'' Universe,
\beq
\frac{d\sigma_{ab}}{d \lambda}
\approx -\frac{1}{l^6} \Sigma_{ac} \Sigma^{c}_{~b}+\frac{1}{l^6} \Sigma^2 \left(g_{ab}+ k_a k_b\right) ~,\label{nullsigmaab}
\eeq
where we left only two $l^{-6}$ terms.
If we substitute Equations (\ref{sigmas}) into (\ref{nullsigmaab}), then we finally get
\bea
\frac{d\sigma_{ab}}{d \lambda}
 &\approx& \frac{ \Sigma^2}{l^6}~ {\rm tr} \left( g_{ab} - 2 g^{ab} \right) ~.\label{finaleqnull}
\eea
By integrating Equation (\ref{finaleqnull}), we have the following approximation on the shear,
\bea
\sigma_{ab} (\lambda) 
 &\approx& \sigma_{ab} (0) + \Sigma^2 \int_0^{\lambda} \frac{1}{l^6}~ {\rm tr}\left( g_{ab} - 2 g^{ab} \right) d \lambda ~.
\eea

\bigskip

\section{~Conclusions}

\setcounter{equation}{0}
\renewcommand{\theequation}{\arabic{section}.\arabic{equation}}

In the standard point-particle inflationary cosmology,
the influence of the shear on the ensuing Universe
evolution are negligible to produce the homogeneous and isotropic Universe features.
It is worthy to note that in the homogeneous and ``anisotropic'' Universe, one can have the condition
$\sigma_{ab} \neq 0$. Here the non-vanishing
$\sigma_{ab}$ evolves and dominates in the early anisotropic Universe.
In the ``extremely early'' Universe, if the scale factor $l(t)$ increases, then the evolution equation for the shear, ${d\sigma_{ab}}/{d t}$, decreases.
This means that the influence of the shear decreases as the scale of the Universe increases.


Recently, the two detectors of the LIGO (Laser Interferometer Gravitational-Wave Observatory) simultaneously observed a transitory gravitational-wave signal which  directly matches the waveform predicted by Einstein's general relativity for the inspiral and merger of a pair of stellar-mass black holes and the ringdown of the resulting single black hole \cite{LIGOPRL16}.
In the source frame, the initial black hole masses are $36 M_{\bigodot}$ and $29 M_{\bigodot}$, and the final black hole mass is $62 M_{\bigodot}$, with $3.0  M_{\bigodot} c^2$ radiated in gravitational waves  \cite{LIGOAPJL16}.
This detection is the first step to discovery of the gravitational wave background (GWB) radiation. Unfortunately, the LIGO's detection sensitivity at low frequencies is limited by the largest practical arm lengths, by terrestrial gravity gradient noise, and by interference from nearby moving objects. Future gravitational wave observatories like the Evolved Laser Interferometer Space Antenna (eLISA), might show ``primordial'' gravitational waves generated during cosmological inflation, relics of the early Universe, up to less than a second of the Big Bang \cite{eLISA2012}.
If we can observe any anisotropy of the ``primordial'' gravitational wave background radiation, then the evolution equation for the shear in this article will be responsible to the future observational anticipations for the gravitational waves.

 There are many possibilities to develop our model of the early anisotropic Universe. For instance, we may further consider the rotational motions of the early Universe because of the initial conditions of the Bianchi type I Universe.
Instead of using the Bianchi type I, we may consider other anisotropic Universe models that might describe the rotational motions.
Applying string theory \cite{Polchinski98, GSW87} to anisotropic Universe models wolud further lead us to consider the rotational motions and geodesic surface congruence in the early Universe.
There are previous results about the anisotropic Universe with cosmic strings and bulk viscosity (e.g., \cite{TNSR2008, TNSR2009a, TNSR2009b, TBR2010, PC2011}) and Bianchi type I string cosmological model in general relativity.

\begin{acknowledgements}
We would like to thank Prof. Yong Seung Cho for his suggestions and comments.
\end{acknowledgements}


\begin{thebibliography}{99}

\bibitem[{LIGO PRL} 2016]{LIGOPRL16}
B. P. Abbott et al. (LIGO Collaboration), 
Phys. Rev. Lett. {\bf 116}, 061102 (2016)

\bibitem[{LIGO APJL} 2016]{LIGOAPJL16}
B. P. Abbott et al. (LIGO Collaboration), 
Astrophys. J. Lett. {\bf 818}:L22 (2016)


\bibitem[{Adhav et al.} 2011]{ABDR2011} K. S. Adhav,  A. S. Bansod, M. S. Desale,  R. B.Raut, Astrophys. Space Sci. {\bf 331}, 689 (2011)

\bibitem[{Adhav et al.} 2010]{AGKB2010}  K. S.Adhav,  P. S.Gadodia,  S. D.Katore,   A. S.Bansod, Astrophys. Space Sci. {\bf 327}, 125 (2010)

\bibitem[{eLISA} 2012]{eLISA2012} P. Amaro-Seoane et al. (2012), arXiv:1201.3621 


\bibitem[{Byland and Scialom} 1998]{BS98}  S.Byland,   D.Scialom, Phys. Rev. D {\bf 57}, 6065 (1998)

\bibitem[{C\'aceres et al.} 2010]{Caceres2010} D. L. C\'aceres, L. Castaneda,  J. M.Tejeiro, ``Shear dynamics in Bianchi I cosmology'', arXiv:1003.3491 [gr-qc] (2010)

\bibitem[{Carroll} 2004]{Carroll2004} S. M.Carroll,  An Introduction to General Relativity:
Spacetime and Geometry. Addison-Wesley, New York (2004)




\bibitem[{Chiba et al.} 1997]{Chiba97}T. Chiba et al., Phys. Lett. B {\bf 408}, 47 (1997)


\bibitem[{Cho and Speliotopoulos}
1995]{CS95}  H. T.Cho, A. D. Speliotopoulos, Phys. Rev. D {\bf 52}, 5445 (1995)

\bibitem[{Cho and Hong} 2007]{CH2007} Y. S. Cho,  S. T. Hong,  Phys. Rev. D {\bf 75}, 127902 (2007)

\bibitem[{Cho and Hong} 2008]{CH2008} Y. S. Cho ,  S. T. Hong, Phys. Rev. D {\bf 78}, 067301 (2008)

\bibitem[{Cho and Hong} 2011]{CH2011} Y. S. Cho , S. T. Hong,  Phys. Rev. D {\bf 83}, 104040 (2011)


\bibitem[{Cho and Hong (An extract from a book)} 2011]{CH2011Book} Y. S. Cho ,  S. T. Hong,  in Aspects of Today's Cosmology, ed. Antonio Alfonso-Faus (Rijeka: InTech) {\bf 347}  (2011)

\bibitem[{Ellis} 2009]{Ellis2009} G. F. R. Ellis,  Relativistic cosmology. Gen. Relativ. Gravit. {\bf 41}, 581 (2009)



\bibitem[{Ellis and van Elst} 1999]{EV99}  G. F. R. Ellis, H. van Elst, Cosmological models (Carg\'ese lectures 1998). NATO Sci. Ser. C: Math. and Phys. Sci. {\bf 541}, 1-116 (1999)





\bibitem[{Green et al.} 1987]{GSW87} M. B. Green,  J. H. Schwarz,  E.Witten,  Superstring Theory. Cambridge Univ. Press, Cambridge  (1987)





\bibitem[{Hawking and Ellis} 1973]{HE73}  S. W. Hawking,   G. F. R.Ellis, The Large Scale Structure of Space-Time.  Cambridge Univ. Press, Cambridge  (1973)

\bibitem[{Hawking and Penrose} 1970]{HP70}  S. W.Hawking,  R.Penrose,  Proc. Roy. Soc. Lond. A {\bf 314}, 529  (1970)



\bibitem[{Jacobs} 1969]{Jacobs69}  K. C. Jacobs,  Bianchi type I cosmological models. Dissertation (Ph.D.), California Institute of Technology (1969)



\bibitem[{Jamil et al.} 2012]{JAMM2012}  M. Jamil, S. Ali, D. Momeni, R. Myrzakulov, Eur. Phys. J. C {\bf 72}:1998 (2012)


\bibitem[{Kasner} 1921]{Kasner21} E. Kasner, in American Journal of Mathematics (Vol. {\bf 43}, No. 2), p. 217-221 (1921)







\bibitem[{Misner et al.} 1973]{MTW} C. W. Misner, K. P. Thorne, J. A. Wheeler, Gravitation. Freeman, San Francisco (1973)




\bibitem[{Pacif and Mishra} 2015]{PM2015}  S. K. J.Pacif,  B.Mishra, Astrophys. Space Sci. {\bf 360}, 48 (2015)



\bibitem[{Pradhan and Chouhan} 2011]{PC2011}  A.Pradhan,  D. S.Chouhan, Astrophys. Space Sci. {\bf 331}, 697 (2011)


\bibitem[{Polchinski} 1998]{Polchinski98}  J.Polchinski, String Theory. Cambridge Univ. Press, Cambridge (1998)


\bibitem[{Saha} 2007]{Saha2007}  B.Saha, Astrophys. Space Sci. {\bf 312}, 3 (2007)

\bibitem[{Sharif and Zubair} 2010]{SZ2010}  M. Sharif,  M.Zubair, Astrophys. Space Sci. {\bf 330}, 399 (2010)

\bibitem[{Sharif and Saleem} 2015]{SS2015} M.Sharif,   R.Saleem, Astrophys. Space Sci. {\bf 360}, 46 (2015)

\bibitem[{Shamir} 2015]{Shamir2015} M. F. Shamir,  Eur. Phys. J. C {\bf 75}:354 (2015)



\bibitem[{Singh and Kale} 2011]{SK2011}  G. P. G.Singh,  A. Y.Kale, Astrophys. Space Sci. {\bf 331}, 207 (2011)


\bibitem[{Singh and Bishi} 2015]{SB2015}  G. P. G.Singh,  B. K.Bishi, Astrophys. Space Sci. {\bf 360}, 34 (2015)



\bibitem[{Tiwari} 2008]{Tiwari2008}  R. K. Tiwari, Astrophys. Space Sci. {\bf 318}, 243 (2008)


\bibitem[{Tripathy et al.} 2008]{TNSR2008}   S. K. Tripathy,  S. K. Nayak, S. K. Sahu,   T. R. Routray, Astrophys. Space Sci. {\bf 318}, 125 (2008)


\bibitem[{Tripathy et al. 2009a}]{TNSR2009a}  S. K. Tripathy,  S. K.Nayak,  S. K. Sahu,  T. R. Routray, Astrophys. Space Sci. {\bf 321}, 247 (2009a)


\bibitem[{Tripathy et al. 2009b}]{TNSR2009b}  S. K. Tripathy,  S. K.Nayak,  S. K.Sahu,  T. R.Routray, Astrophys. Space Sci. {\bf 323}, 91 (2009b)



\bibitem[{Tripathy et al.} 2010]{TBR2010}  S. K.Tripathy,  D.Behera,  T. R. Routray, Astrophys. Space Sci. {\bf 325}, 93 (2010)


\bibitem[{Tsagas et al.} 2008]{Tsagas2008} C. G.Tsagas  et al.: Phys. Rep. {\bf 465}, 61 (2008)

\bibitem[{Wainwright and Ellis} 1997]{WE97}  J. Wainwright,  G. F. R. Ellis, Dynamical Systems in Cosmology. Cambridge Univ. Press, Cambridge (1997)



\bibitem[{Wald} 1984]{Wald84}  R. M. Wald, General Relativity. The Univ. of Chicago Press, Chicago (1984)








\end{thebibliography}
\end{document}